# Quantifying resilience to recurrent ecosystem disturbances using flow-kick dynamics

Katherine Meyer[1,]*, Alanna Hoyer-Leitzel[2], Sarah Iams[3], Ian Klasky[4,5], Victoria Lee[4], Stephen Ligtenberg[4], Erika Bussmann[6], and Mary Lou Zeeman[4]

[1]*University of Minnesota Department of Mathematics, Minneapolis, MN;*
[2]*Mount Holyoke College Department of Mathematics and Statistics, South Hadley, MA;*
[3]*Harvard University John A. Paulson School of Engineering and Applied Sciences, Cambridge, MA;*
[4]*Bowdoin College Mathematics Department, Brunswick, ME;*
[5]*University of Colorado Department of Applied Mathemtaics, Boulder, CO;*
[6]*Minnetonka High School, Minnetonka, MN*
*Corresponding author*: `meye2098@umn.edu`

Shifting ecosystem disturbance patterns due to climate change (e.g. storms, droughts, wildfires) or direct human interference (e.g. harvests, nutrient loading) highlight the importance of quantifying and strengthening the resilience of desired ecological regimes. Although existing metrics capture resilience to isolated shocks, gradual parameter changes, and continuous noise, quantifying resilience to repeated, discrete disturbances requires novel analytical tools. Here we introduce a *flow-kick* framework that quantifies resilience to disturbances explicitly in terms of their magnitude and frequency. We present a *resilience boundary* between disturbances that cause either escape from a basin of attraction or stabilization within it, and use the resilience boundary to build resilience metrics tailored to repeated, discrete disturbances. The flow-kick model suggests that the distance-to-threshold resilience metric overestimates resilience in the context of repeated disturbances. It also reveals counterintuitive triggers for regime shifts, such as increasing recovery times between disturbances, or increasing disturbance magnitude and recovery times proportionately.

## Introduction

Climate change and other human impacts are altering disturbance patterns in Earth's systems. Shifting patterns of precipitation[1], drought[2], fires[1,2,3], harvests[4], and nutrient loading[4], coupled with society's dependence on healthy ecosystems[5], underscore the need to quantify and enhance the resilience of desired ecological regimes[6,7]. Resilience is commonly defined as a system's capacity to absorb change and disturbance while maintaining its structure and function[6]. Translating this qualitative definition to quantitative metrics for resilience requires clarity regarding both the system properties to be preserved ("resilience *of* what") and the types of disturbances under consideration ("resilience *to* what")[7].

Existing resilience metrics typically measure resilience of a basin of attraction, and each captures resilience for a specific disturbance type[8] (Table 1). For example, the width or volume of an attractor's basin[9,10,11] or the distance from an attractor to a threshold in state space[12] reflect resilience to potentially large, isolated disturbances (e.g. 100-year flood). The weakest eigenvalue of the linear approximation to a system near an attracting equilibrium can indicate recovery rates after a small, isolated disturbance[13] (e.g. minor drought), while expected escape time under a diffusion process quantifies resilience to continual random perturbation (e.g. environmental stochasticity)[14]. Lastly, the distance to a bifurcation in parameter space[15] indicates resilience to gradual changes in an environmental parameter (e.g. rising carbon dioxide concentration).

This article is about the fourth row in Table 1, which introduces new metrics to quantify resilience of ecosystems that are shaped by discrete, repeated disturbances ("kicks"). In some cases, nature provides these kicks: hurricanes deliver repeated blows to reefs, damaging coral and displacing herbivores that keep macroalgae in check[16], wildfires repeatedly combust tracts of biomass in grassy/woody ecosystems[2,17,18], and extreme precipitation events or droughts impact both natural and human communities. In other cases, humans deliver kicks directly via unintentional impacts (e.g. agricultural nutrient runoff) and by intentional management (e.g. prescribed burns[3], introduction of predators for pest management[19], or harvests).

When the timescales of the kicks' recurrence and the system's recovery coincide, specific disturbance characterisitcs[20] such as kick magnitude and frequency drive the outcomes. For example, different intensities and frequencies of fire events in the North American Great Plains promote dominance of grassland, shrubland, or woodland[17], while in reef ecosystems, the frequency of hurricanes influences coral-macroalgae competitive outcomes[16]. Here we describe a new *flow-kick* framework, specially equipped to quantify resilience of a basin of attraction to repeated, discrete disturbances. Using an example from fisheries, we show the shortcomings of resilience metrics based in state space for detecting this type of resilience, introduce the flow-kick model of disturbance, and propose new resilience metrics based in what we term disturbance space. In the context of a lake eutrophication model, we describe connections between this new framework and existing resilience metrics, and generalize the approach to include stochastic kicks and recovery times. Lastly, we use a model from Earth's climate system to highlight counterintuitive behaviors in flow-kick systems with more than one dynamic variable.

**Table 1 | Summary of resilience metrics**

| | Measurement context | Metric diagram | Metric description | Disturbance type |
|---|---|---|---|---|
| Existing metrics | State space | | a. basin width / volume[9,10,11] | isolated perturbation |
| | | | b. distance to threshold[12] | Isolated perturbation |
| | Time | | c. characteristic return time[13] (from eigenvalues) | small, isolated perturbation |
| | | | d. expected escape time[14] (stochastic diffusion) | continuous, noisy perturbations |
| | Parameter space | | e. distance from parameter value to bifurcation value[15] | gradual environmental change |
| New metrics | Disturbance space | | f. area above resilience boundary | repeated, discrete disturbances |
| | | | g. area between disturbance and resilience boundary | disturbance in addition to repeated, discrete pattern |

# Flow-kick based resilience quantification in a fishery example

**A deterministic fish population model.** Instances of fishery collapse align conceptually with population models that include an Allee threshold, a critical stock size above which an undisturbed population grows to carrying capacity and below which the population collapses[21,22,23]. We use a minimal model of the Allee effect[21] (Methods Section 1) to represent the dynamics of an undisturbed fish population. Figure 1a shows population growth rates as a function of population size for two hypothetical fisheries. Each population has an attracting carrying capacity of 100 kilotonnes (kt) and a repelling Allee threshold at 20 kt. The latter constitutes the lower boundary of the carrying capacity's basin of attraction, which is highlighted in yellow. How resilient is this basin is to repeated (instantaneous) harvests? Two common resilience metrics fail to distinguish between populations 1 and 2. The distance-to-threshold metric (here, the distance from the carrying capacity to the Allee threshold) gives the same number (80 kt) for both populations. Similarly, return rates based on linearization about the equilibrium at carrying capacity are identical for the two populations, since the growth rate curves are tangent at this equilibrium. However, one might expect population 1 to be more resilient to repeated harvests, since its growth rate exceeds that of population 2 at all abundances between 20 kt and 100 kt. These growth rates determine how transient recovery dynamics will balance (or fail to balance) repeated disturbances, and are not captured by distance-to-threshold or linearized return rate metrics.

**Incorporating kicks and flows.** To devise a resilience metric that can distinguish populations 1 and 2, we represent harvests as kicks that periodically and instantaneously decrease the population size by a certain amount. During the period between kicks, the population increases or decreases according to the model for the undisturbed system (Figure 1a and Methods Section 1). We call this fixed period between kicks the flow time, and refer to the new dynamics built from repeatedly alternating flows and kicks as a flow-kick system. Figure 1b contrasts two flow-kick trajectories for population 1 that each start at carrying capacity. Trajectory $S$ results from flow times of 3 months (solid lines) and kicks of -12 kt (dashed lines), representing a quarterly 12 kt harvest. This disturbance causes the population to stabilize in oscillations between roughly 73 kt (post-kick) and 85 kt (pre-kick). Trajectory *C* results from a flow time of 10 months and a kick of -40 kt; this combination drives the population below the Allee threshold, beyond which the kicks merely hasten collapse. We say that the basin of attraction is resilient to the first disturbance but not to the second. Suprisingly, the average harvest rate is the same (48 kt/yr) for both trajectories. A natural question arises: which disturbance patterns, like *C*, drive the population out of the basin of attraction, and which, like *S*, allow the population to stabilize within the basin?

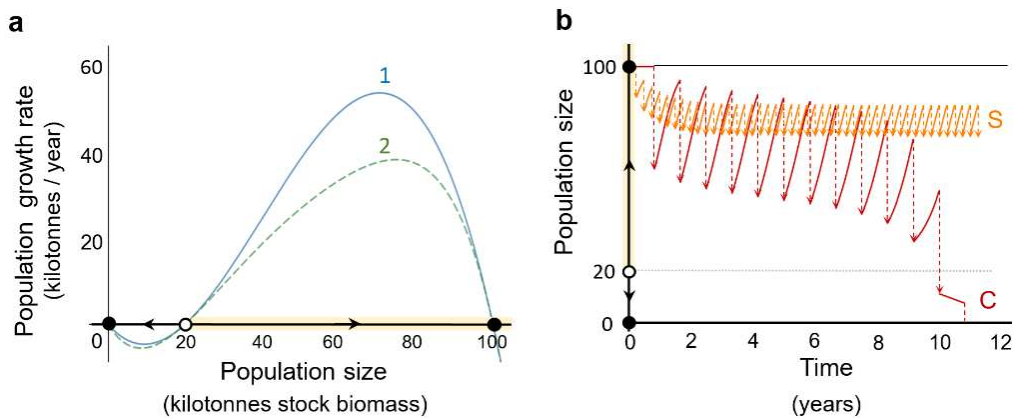

**Figure 1 | A flow-kick model of repeated harvests from a fishery. a**, Growth rate curves for two fish populations (1 and 2), which differ in growth strength but each have an attracting equilibrium (carrying capacity) at 100 kilotonnes (kt) and a repelling equilibrium (Allee threshold) at 20 kt. Arrows of the phase-line diagram superimposed on the horizontal axis show the direction of population growth or decline; the basin of attraction of the carrying capacity equilibrium is highlighted in yellow. **b**, Trajectories of population 1 under two harvesting patterns: $S$, 12 kt harvested every 3 months stabilizes between roughly 73 kt (post-kick) and 85 kt (pre-kick) and $C$, 40 kt harvested every 10 months, collapses. Note that both harvesting patterns remove fish at the same average rate. Dashed arrows show instantaneous harvests (kicks) and solid curves show recovery (flow).

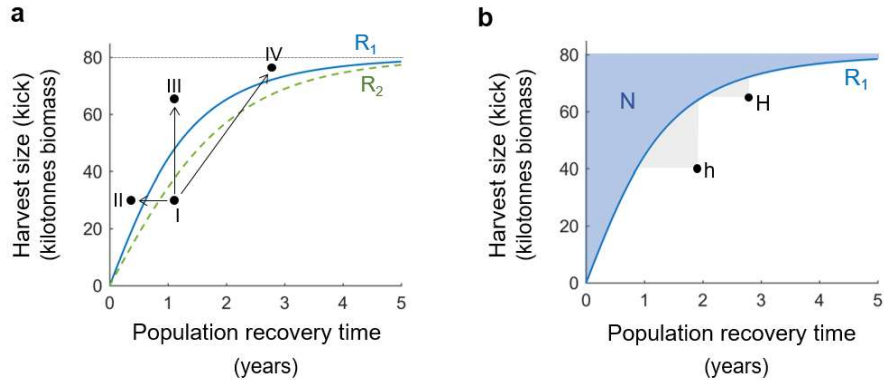

**Figure 2 | The resilience boundary in disturbance space. a**, Disturbance space represents disturbances by recovery time (horizontal axis), and kick size (vertical axis). Harvest patterns above the resilience boundary $R_1$ cause population 1 to collapse, while those below it allow stabilization within the basin of attraction of carrying capacity. The position of population 2's resilience boundary, $R_2$, below $R_1$ indicates relatively lower resilience to harvests. Arrows highlight three triggers for population collapses: decreased recovery time (I→II), increased harvest amount (I→III), and proportionate increases in recovery and harvest (I→IV). **b**, Resilience metrics based in disturbance space include the area of the non-resilient region above a resilience boundary (N) and the area between a harvesting strategy and the resilience boundary (light grey).

**The resilience boundary.** We visualize the answer in disturbance space, where the horizontal axis gives flow time and the vertical axis gives kick magnitude (Figure 2a and Methods Sections 2, 3). Each point in disturbance space represents a specific disturbance pattern; those above the curve $R_1$ cause population 1 to collapse, while those below $R_1$ lead to stabilization within the basin of attraction of the carrying capacity. We call $R_1$ the resilience boundary for population 1. As recovery times increase, $R_1$ approaches a horizontal asymptote at 80 kt, the distance from the attracting equilibrium to the Allee threshold. The concave shape of $R_1$ means that a regime shift from a viable to a collapsed fishery can be triggered not only by decreases in flow time (I→ II) or increases in kick size (I→ III) but also by simultaneous increases in flow time and kick sizes that maintain their ratio (i.e. the average harvest rate, as in Figure 1b) (I→IV).

The curve $R_2$ in Figure 2a is the resilience boundary for population 2. It lies strictly below $R_1$, so for any given recovery time, population 1 can withstand larger harvests than population 2. Equivalently, any given harvest size can be taken more frequently from population 2 than from 1 without collapsing the population. The area between $R_1$ and $R_2$ consists of disturbances to which population 1 is resilient but population 2 is not.

**Quantifying intrinsic resilience.** Resilience boundaries in disturbance space provide new options for quantifying resilience to repeated perturbations. At first glance, one might be inclined to calculate the area under the resilience boundary; however, this area is infinite. On the other hand, the area of the region between a resilience boundary, its horizontal asymptote, and the vertical axis (dark shaded region, N, in Figure 2b) is finite for most growth rate curves (Supplementary Information Section 2). The area of N measures the collection of disturbance patterns (with kick magnitude less than the distance to threshold) to which a population is not resilient. Since it does not presume any baseline disturbance pattern, one might call it a measure of intrinsic non-resilience. By this method, Fisheries 1 and 2 have intrinsic non-resiliences of approximately 99 kt-yr and 127 kt-yr, respectively (Methods Section 3). More meaningful than the numbers alone may be their difference, 28 kt-yr, representing the area of the region between $R_1$ and $R_2$.

**Quantifying resilience of a management strategy.** A fishery manager who has selected a harvesting pattern (e.g. point *h* in Figure 2b) may be concerned with the resilience of the population to disruptions beyond the harvests. The horizontal and vertical distances from *h* to $R_1$ measure the resilience of *h* to increased kick frequency and kick size, respectively, while the area of the lightly shaded region extending from the point *h* to the resilience boundary provides a metric of the overall resilience built into that management strategy. By all these metrics, the harvesting pattern *H* has lower resilience than *h*.

**Revisiting the distance-to-threshold concept.** The distance from an attracting system state to the boundary of its basin of attraction in an undisturbed system overestimates resilience to an external shock in the context of accumulating repeated disturbances. To illustrate, Figure 3 shows more detailed dynamics when 12 kt are harvested from population 1 every 3 months. We use the potential function $V_1$ to visualize the flow-kick dynamics. We imagine the system's (undisturbed) dynamics as a ball rolling downhill along $V_1$ at a speed proportional to the slope. The horizontal position of the ball encodes the population size. The attracting equilibria at 0 kt and 100 kt appear at the bottoms of their respective basins of attraction, while the repelling equilibrium at 20 kt separates these basins at a peak.

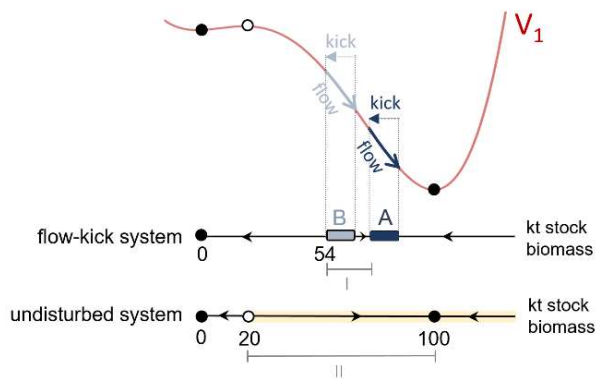

**Figure 3 | Shorter distance-to-threshold in a flow-kick system.** The potential function $V_1$ encodes dynamics of the undisturbed fish population 1 (c.f. Figure 1a). Intervals over which kick and flow (harvest of 12 kt every 3 months) balance are superimposed on $V_1$ and depicted in a phase-line diagram for the flow-kick system, below. The distance I between attracting equilibrium interval *A* and repelling (threshold) equilibrium interval *B* represents the largest external shock that the flow-kick system can absorb. It is much smaller than the distance II from the carrying capacity to the Allee threshold of the undisturbed system.

We depict flows along the surface of $V_1$ and show kicks as horizontal translations. The flow and kick balance at exactly two intervals within the desired basin of attraction. The first flow-kick equilibrium interval, *A*, attracts flow-kick trajectories and confers resilience to the disturbance; a flow-kick trajectory that starts at carrying capacity stabilizes at this attracting interval as in Figure 1b. However, the second flow-kick equilibrium interval, *B*, repels flow-kick trajectories, so those that start below 54 kt will collapse to zero. If population 1 was managed at the attracting flow-kick equilibrium interval *A* by harvesting 12 kt stock every 3 months, an external shock that decreased the population by just 35 kt would bring the population below 54 kt, leading to a population collapse if the harvesting strategy continued. Note that such a fatal shock need not push the population below the Allee threshold (20 kt). Thus, the relevant threshold in the context of repeated disturbances is not the basin boundary for the undisturbed system, but a new boundary brought about by flow-kick dynamics. The distance from an attracting flow-kick equilibrium to a flow-kick threshold (I in Figure 3) could provide an alternative metric for resilience of a management strategy. (See the Supplementary Information, Section 7 for a description of the resilience boundary as a bifurcation curve at which the attracting and repelling flow-kick equilibrium intervals coalesce.)

## Nutrient pulses, lake eutrophication, and relationships between resilience metrics

The flow-kick framework can be used to model a wide variety of systems subject to repeated disturbances whose effects transpire rapidly relative to the dynamics of the undisturbed system. In this section, we use an example of lake water quality to illustrate connections between flow-kick and existing resilience metrics.

**A model of lake water quality.** We use a minimal model of alternative stable states to represent water quality in a lake prone to eutrophication[24,25] (Methods Section 1). In the absence of strong nutrient pulses (kicks), the amount of phosphorus (P) in the lake water changes according to the balance between inputs and losses (Figure 4a). P inputs have a sigmoidal shape (purple curve), due to the nonlinear response of nutrient recycling from sediments as the level of P in the lake water increases. Another input, the steady background rate of P contribution from the watershed, determines the height of this S-shaped curve. Simultaneously, the lake loses P via outflow and/or sedimentation at a rate proportional to the level of P (green line). At the intermediate level of background watershed P input shown in Figure 4a, multiple equilibrium levels of P occur at the intersections between the input and loss curves. For the parameters we have chosen (Methods Section 1), a low-P attracting equilibrium at ~50 hectograms (hg) corresponds to an oligotrophic lake—a desirable regime with clear water, low algae levels, and healthy fish populations. A high-P attracting equilibrium at ~145 hg corresponds to a eutrophic lake—an undesirable regime with turbid water, algal overgrowth, and oxygen depletion that kills fish. Feedbacks between the biology and turbidity further stabilize the two regimes. A repelling equilibrium at intermediate P levels (100 hg) separates the two alternate stable regimes. The arrows on the horizontal axis in Figure 4a summarize the dynamics. The preferred, oligotrophic basin of attraction in this system is a literal basin of the potential function $V_2$, and is highlighted in yellow on the horizontal axis.

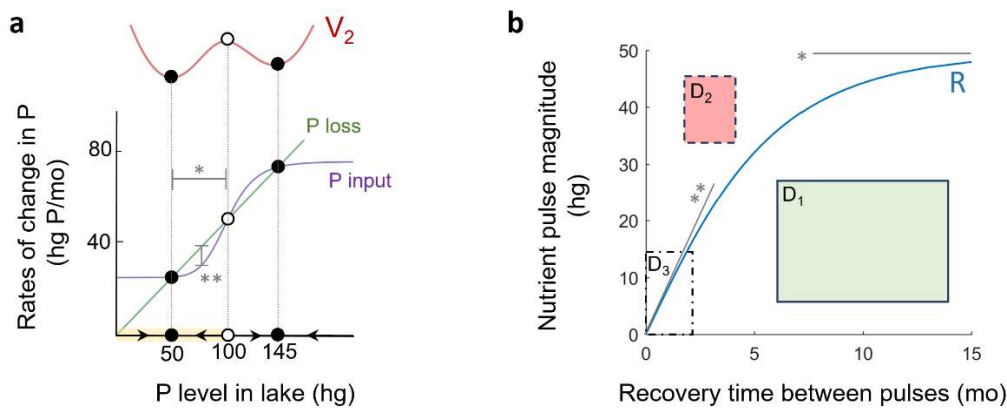

**Figure 4 | Connections between the resilience boundary and existing resilience metrics illustrated in a model of lake water quality. a**, Phosphorus (P) dynamics in a model lake summarized by a potential function ($V_2$), loss and input rates (green line and purple curve), and a phase-line diagram (horizontal axis). The basin of attraction for the desired oligotrophic equilibrium at 50 hectograms (hg) P is highlighted in yellow; a repelling equilibrium at 100 hg P separates this basin from that of the attracting eutrophic equilibrium at 145 hg P. **b**, The resilience boundary (R) for the desired basin features a horizontal asymptote (*) that matches the distance to threshold in state space (* in **a**) and a limiting slope near the origin (**) that matches the distance to bifurcation in parameter space (** in **a**) The oligotrophic basin is or is not resilient to stochastic flow-kick disturbances drawn, respectively, from a rectangle below R (e.g. $D_1$) or a rectangle above R (e.g. $D_2$). The flow-kick framework does not currently predict outcomes of stochastic disturbances drawn from either side of R (e.g. $D_3$).

In addition to the background level of P input from the lake's watershed, we represent pulses of nutrient input as kicks that increase P. Such pulses could come, for example, from precipitation events following fertilizer applications to fields in the watershed. As before, we ask which magnitudes and frequencies of nutrient kicks push the lake from its oligotrophic equilibrium to the eutrophic basin of attraction, and which allow the lake to stabilize within the oligotrophic basin. The resilience boundary $R$ in disturbance space (Figure 4b) separates disturbances to which the oligotrophic basin is resilient (below $R$) or is not (above $R$).

**Bridging existing resilience metrics.** The resilience boundary $R$ simultaneously generalizes the distance-to-threshold and distance-to-bifurcation resilience metrics in Table 1. For large recovery times, $R$ approaches a horizontal asymptote (* in Figure 4b) whose height is the distance from the attracting low P equilibrium (50 hg) to the threshold (100 hg) in state space (* in Figure 4a). This phenomenon also occurred in the fisheries example and is not surprising: kicks just smaller than the distance to the basin boundary can be balanced by long enough recovery times, while kicks larger than this distance inevitably cause escape from the basin of attraction. Less obvious is that the slope of $R$ near the origin (** in Figure 4b) matches the maximum difference between P output and P input between 50 hg and 100 hg (** in Figure 4a). This maximum distance represents the amount by which background nutrient loading could increase without losing a low-P equilibrium. Thus, it is a distance to bifurcation in parameter space, where background nutrient loading is the parameter. The match stems from the fact that as the flow time and the kick decrease toward zero in a fixed ratio (an average nutrient addition rate), the flow-kick system limits to a continuous system with P input continuously augmented by this average rate (Supplementary Information Section 3). These echoes of existing resilience metrics in the resilience boundary $R$ reiterate that distance-to-bifurcation gives a good measure of resilience to nearly continuous disturbance (small kicks and short flow times, such as those in box $D_3$ in Figure 4b), while the distance-to-threshold indicator approximates resilience to very rare disturbances (large flow times). The flow-kick model bridges these metrics, predicting resilience to repeated disturbances of intermediate size and frequency.

**Stochastic disturbance.** Although we use a deterministic framework to build the resilience boundary, it also supports predictions when kicks and flows occur stochastically. If the random flow time and kick size are contained within a bounded rectangular set that lies either entirely below $R$ (e.g. box $D_1$ in Figure 4b) or above $R$ (e.g. box $D_2$ in Figure 4b), then a randomized flow-kick trajectory starting at the low-P equilibrium will either stay within its basin of attraction indefinitely or escape in finite time, respectively. This result holds for any flow-kick system with a single dynamic variable (Supplementary Information Section 4). It does not, however, predict the outcome of randomized flows and kicks that are drawn from either side of $R$, such as from box $D_3$ in Figure 4b.

Our use of bounded sets for the possible values of stochastic kicks and flows differs from stochastic diffusion models of disturbance[14] in which the disturbance occurs continuously in time and can be arbitrarily large in magnitude (with ever smaller probability). In such a system, the question becomes not whether the state will escape a basin of attraction but when, and by what path. In many systems, including lakes, stochastic noise may be an appropriate approximation to disturbance, and the stochastic diffusion approach could help predict outcomes of flows and kicks from both sides of the resilience boundary[26]. However, disturbances in other systems such as storms, fires, and harvests may be bounded by maximum kicks or minimum recovery times. In these cases, the flow-kick framework may align better with the disturbance type.

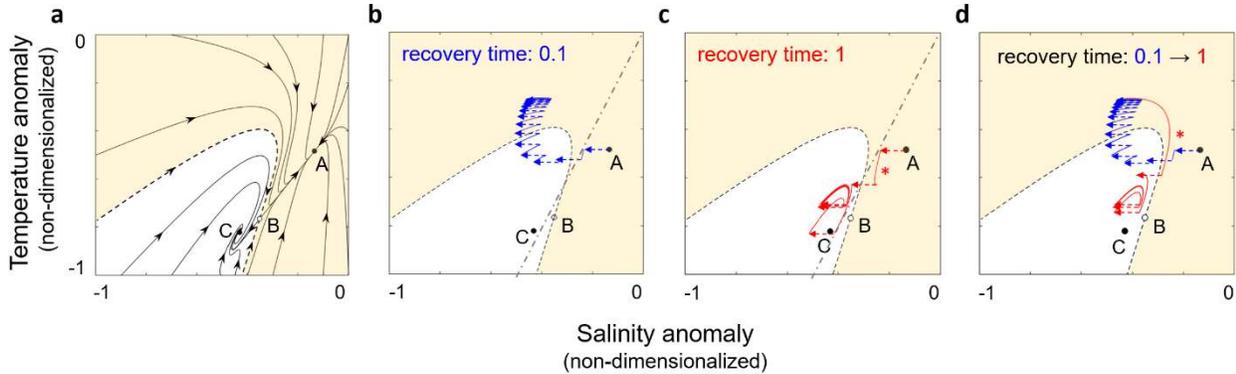

**Figure 5 | Flow-kick dynamics in a two-dimensional model of ocean circulation. a**, Phase portrait for Stommel's ocean box model with two state variables: non-dimensional salinity and temperature anomalies of the higher latitude box (e.g. the North Atlantic). The attracting equilibria *A* and *C* correspond to opposite circulation directions, and the saddle equilibrium *B* lies on the boundary of the desired state *A*'s basin of attraction, shaded yellow. Dashed arrows in **b-d** represent identical freshwater pulses (kicks) from glacial meltwater. For trajectories that start at *A*, relatively shorter recovery times cause stabilization within *A*'s basin (**b**), while longer recovery times lead to escape from *A*'s basin (**c**). Lengthening recovery times can, counterintuitively, trigger escape from a basin (**d**). Asterisks in **c**,**d** highlight flows within *A*'s basin that carry the state away from *A*, a mechanism behind this unexpected phenomenon. The dash-dot line in **b**,**c** separates regions of state space that yield different circulation directions; note they do not align with basins of attraction.

## Further applications

Flow-kick applications extend beyond ecosystems. Our final example, taken from the climate system, illustrates the potential complexity of flow-kick dynamics involving more than one state variable.

**A model of ocean circulation.** We use Stommel's ocean box model[27] (Methods Section 1) as a low-dimensional representation of the Atlantic Meridional Overturning Circulation (AMOC), which currently delivers warm waters to Western Europe. Gradients in seawater temperature, salinity, and hence density drive this circulation. By tracking temperature and salinity in two idealized ocean boxes, Stommel identified two attracting circulation patterns. The first, driven by sinking of cold waters in the box that we interpret as the North Atlantic, corresponds to the direction we see today. The second, driven by the buoyancy of low-salinity North Atlantic waters, circulates in the opposite direction. This alternate circulation pattern in the model can be interpreted as a collapse of the AMOC, which could cool Europe by several degrees Celsius and impact climate on a global scale[28]. Figure 5a gives a phase portrait for the model system. The horizontal and vertical axes represent the two dynamic variables: non-dimensional salinity and temperature, respectively, in the North Atlantic. The attracting equilibrium *A* yields the current circulation direction, while the attracting equilibrium *C* reverses the direction; they are separated by a saddle equilibrium *B*, whose stable manifold (dashed line) forms the boundary between the basins of attraction of *A* and *C*.

**Counterintuitive flow-kick behavior.** In the past, pulses of low-salinity glacial meltwater to the North Atlantic may have perturbed the AMOC from temperature- to salinity- dominated circulation (e.g. from regime *A* to regime *C* in Figure 5). This mechanism is hypothesized to have caused the Younger Dryas, an anomalous cooling in the northern hemisphere around 12,000 years ago[29,30].

We represent such a pulse of meltwater as a kick (-0.1 salinity units, 0 temperature units) that moves the system state to the left (Figure 5b-d). (Note than in systems of multiple dimensions, kicks have magnitude and direction.) When the recovery between kicks lasts 0.1 time units, a trajectory that starts at *A* stabilizes within the basin of attraction of *A* after a temporary excursion through the basin of *C* (Figure 5b). Surprisingly, a longer recovery time of 1 unit causes the trajectory from *A* to stabilize in the basin of attraction of *C* (Figure 5c). This counterintuitive behavior stems from transient dynamics of the undisturbed system within the basin of attraction of *A,* which involve an excursion away from *A* (e.g. flows marked by * in Figure 5c,d) before ultimately returning. The phenomenon in which increasing recovery time between disturbances triggers escape from a basin (Figure 5d) does not occur in models with one state variable, but this climate example alerts us to the possibility of similar behavior in ecological and other systems with multiple state variables.

## Discussion

Flow-kick systems can model disturbance and resilience in diverse settings, from the examples presented here to the resilience of a financial network to repeated shocks[32], or the resilience of an attracting cancer disease state to a series of treatments[33]. We advocate expanding the existing suite of resilience metrics (Table 1) to better measure resilience to repeated, discrete disturbances via flow-kick models. Our proposed metrics, based in disturbance space, capture the dynamic interplay between disturbance and recovery. Though sparse data and incomplete knowledge of specific processes might hinder predictive applications of flow-kick modeling, it may prove useful as a conceptual framework. Previously, low-dimensional models have illuminated the potential for rapid and potentially irreversible changes in ecosystems[24,25,31]. The flow-kick examples we consider highlight underappreciated mechanisms for regime shifts such as proportional increases in the disturbance magnitude and recovery period (Figure 2a, I→IV), flow-kick induced thresholds (Figure 3), and even increases in recovery time with the disturbance magnitude held constant (Figure 5d).

There is another important subtlety in Stommel's box model: it is not true that every point in the basin of attraction of *A* represents the same circulation direction as *A*. Only salinity-temperature combinations to the right of the dash-dot line in Figure 5b,c have that property. This highlights the fact that desirable regions of state space do not necessarily coincide perfectly with basins of attraction. The flow-kick framework can also be used in such cases, to identify disturbance patterns that stabilize a system in a desirable region of state space (MLZ *et al.*, in preparation). An exciting direction for collaborative future research is to combine the flow-kick approach with social science research that identifies desirable properties of state space in the context of sustainability challenges.

## Acknowledgements

This work was supported by an NSF Graduate Research Fellowship (grant number 00039202) to Katherine Meyer, the Mathematics and Climate Change Research Network (NFS grant DMS-0940243), and the Computational Sustainability Network (NSF grant CCS-1521672). Thanks go to Allison Shaw, Lauren Sullivan, Jane Cowles, Kaitlin Kimmel, and Evelyn Strombom for feedback on an early manuscript. We are also indebted to the Ecology Theory Group at the University of Minnesota, and to Ariella Helfgott, Steven Lord, Dick McGehee, and Chris Chong for helpful conversations.

## Author Contributions

MLZ and SI conceived of the flow-kick model of disturbance. All authors contributed to development of the model and analysis of the resulting dynamics. KM drafted the manuscript and SI, AHL, IK, VL, EB, and MLZ contributed to revisions.

## Methods

**1. Models of undisturbed systems.** Our models of undisturbed dynamics consist of ordinary differential equations (ODEs). We use a simple Allee effect model[21] for fish population 1,

$$\frac{dx_1}{dt} = x_1\left(1 - \frac{x_1}{K}\right)\left(\frac{x_1}{A} - 1\right), \quad (1)$$

where $x_1$ is stock biomass (kilotonnes, kt), $t$ is time (years), $K=100$ kt is the carrying capacity, and $A=20$ kt is the critical Allee threshold. We modify the growth rate function (1) for fishery 2 with a factor that reduces its magnitude but preserves its equilibria as well as the slope at the Allee threshold and carrying capacity:

$$\frac{dx_2}{dt} = x_2(1 - \frac{x_2}{100})(\frac{x_2}{20} - 1)(0.0002 x_2^2 - 0.024 x_2 + 1.4) \quad (2)$$

We represent lake phosphorus (P) dynamics with a minimal model of alternative stable states[24,25]

$$\frac{dx}{dt} = l - sx + \frac{rx^q}{m^q + x^q}, \quad (3)$$

where $x$ is the mass of P in the lake (hectograms, hg), $t$ is time (months), $l=25$ is the background watershed input rate (hg/mo), $s=0.5$ mo$^{-1}$ parametrizes P loss, $r=50$ hg/mo is the maximum recycling rate from sediments, $q=8$ parametrizes the shape of the sigmoid input curve, and $m=100$ hg is the half-saturation constant for recycling. The numeric values of these parameters match figure six of Carpenter and colleagues[25], but we modify mass units from kg to hg and time units from years to months.

The dynamics of non-dimensional salinity ($x$) and temperature ($y$) in Stommel's ocean box model[27] are given by

$$\frac{dx}{dt} = \delta(1-x) - \frac{x}{\lambda}|-y + Rx| \quad \text{and} \quad (4)$$

$$\frac{dy}{dt} = 1 - y - \frac{y}{\lambda}|-y + Rx|, \quad (5)$$

where $\delta=1/6$, $\lambda=1/5$, and $R=2$. The variables $x$ and $y$ correspond to the warmer and saltier box in Stommel's system; due to symmetry in the system, $-x$ and $-y$ give salinity and temperature anomalies in the cooler and less saline box that we interpret as the North Atlantic.

**2. Flow-kick model.** We represent the flow-kick process as a discrete map on state space $\mathbb{R}^n$, parameterized by recovery time $\tau$ and kick size $\kappa$. ($\kappa$ is in general an $n$ dimensional vector; in the special case of one state variable, it is simply a number.) Let $\varphi_t(x)$ be the flow function corresponding to an ordinary differential equation for an undisturbed system, giving the position after time $t$ of a trajectory that starts at $x$[34]. Then the map

$$G_{\tau,\kappa}(x) = \varphi_\tau(x) + \kappa \quad (6)$$

represents one cycle of flow-kick, while iteration of $G_{\tau,\kappa}$ represents recurrent disturbances. Disturbance space refers to the $\tau,\kappa$ plane. Fixed points $x^*$ of the flow-kick map occur where the flow and kick exactly balance each other, so $\varphi_\tau(x^*)+\kappa = x^*$. In a one-dimensional system, we call the interval between $x^*$ and the point $x^* - \kappa$ a *flow-kick equilibrium interval*; trajectories flow from $x^*$ to $x^* - \kappa$ and are then kicked back to $x^*$. Stability of equilibria can be determined based on linearization of the flow-kick map (Supplementary Information Section 6).

**3. Numerics and simulations.** We use MATLAB version R2016b to simulate flow-kick trajectories, find flow-kick equilibria, and compute resilience boundaries and the areas they bound. Code mentioned below is available on request. The script *AlleeTrajectories.m* (which calls function *dxdtAllee.m*) was used to simulate flow-kick trajectories for population 1 in Figure 1b, with flow time $\tau = 0.25$ and kick $\kappa = -12$ (trajectory S) and flow time $\tau = 5/6$ and kick $\kappa = -40$ (trajectory C). The MATLAB code *AlleeResBoundary.m* was used to plot the resilience boundaries $R_1$ and $R_2$ in Figure 2 by finding the smallest time needed for the population to recover from a given harvest (see Supplementary Information Section 1). We approximate areas between the resilience boundaries and horizontal asymptotes using numerical integration (trapezoid rule) on a finite interval (script *ResilienceArea.m*), and an analytic bound on the tail from a quadratic growth function (see Supplementary Information Section 2). This non-resilient area metric can be normalized by non-dimensionalising the kick size so the horizontal asymptote is at 1. We calculate the flow-kick equilibria in Figure 3 using the functions *Newton.m* and *CoupledVar.m*, which implement Newton's Method to find zeros of $F(x) = G_{\tau,\kappa}(x) - x$ (see Supplementary Information Section 5). The MATLAB script *LakeResBoundary.m* was used to plot the resilience boundary $R$ in Figure 4 by finding minimum recovery times as in *AlleeResBoundary.m* (see Supplementary Information Section 1). The script *StommelFigure.m* (which calls *dydtStommel.m* and *dydtMinusStommel.m*) was used to create the phase portrait in Figure 5a and to simulate the flow-kick trajectories in Figure 5b-d.

**References**

1. Pachauri, R.K. & Meyer, L.A. (eds.) *Climate Change 2014: Synthesis Report. Contribution of Working Groups I, II and III to the Fifth Assessment Report of the Intergovernmental Panel on Climate Change.* IPCC, Geneva, Switzerland (2014).

2. Dale, V.H. *et al.* Climate change and forest disturbances. *BioScience* **51**, 723–734 (2001).

3. Russell-Smith, J. & Thornton, R. Perspectives on prescribed burning. *Frontiers in Ecology and the Environment* **11**, e3 (2013).

4. Reid, W.V. et al. *Ecosystems and Human Well-being: Synthesis*. Millennium Ecosystem Assessment, Island Press, Washington, D.C. (2005).

5. Kareiva, P., Tallis, H., Rickets, T.H., Daily, G.C & Polasky, S, eds. *Natural Capital: Theory and Practice of Mapping Ecosystem Services*. Oxford University Press, Oxford (2011).

6. Walker, B. & Salt, D. *Resilience Thinking: Sustaining Ecosystems and People in a Changing World.* Island Press, Washington, D.C. (2006).

7. Carpenter, S.R., Walker, B., Anderies, J.M. & Abel, N. From metaphor to measurement: resilience of what to what? *Ecosystems* **4**, 765–781 (2001).

8. Meyer, K. A mathematical review of resilience in ecology. *Nat. Resour. Model.* **29**, 339–352 (2016).

9. Levin, S.A. & Lubchenco, J. Resilience, robustness, and marine ecosystem-based management. *BioScience* **58**, 27–32 (2008).

10. Walker, B., Holling, C.S., Carpenter, S.R. & Kinzig, A. Resilience, adaptability and transformability in social-ecological systems. *Ecol. Soc.* http://www.ecologyandsociety.org/vol9/iss2/art5 (2004).

11. Menck, P.J., Heitzig, J., Marwan, N. & Kurths, J. How basin stability complements the linear-stability paradigm. *Nature Physics* **9**, 89–92 (2013).

12. Beisner, B.E., Dent, C.L. & Carpenter, S.R. Variability of lakes on the landscape: roles of phosphorus, food webs, and dissolved organic carbon. *Ecology* **84**, 1563–1575 (2003).

13. Pimm, S.L. The complexity and stability of ecosystems. *Nature* **307**, 321–326 (1984).

14. Dennis, B., Assas, L., Elaydi, S., Kwessi, E. & Livadiotis, G. Allee effects and resilience in stochastic populations. *Theor. Ecol.* **9**, 323–335 (2015).

15. Ives, A.R. & Carpenter, S.R. Stability and diversity of ecosystems. *Science* **317**, 58–62 (2007).

16. Ippolito, S. & Naudot, V. Alternative stable states, coral reefs, and smooth dynamics with a

kick. *Bull. Math Biol.* 10.1007/s11538-016-0148-2 (2016).

17. Ratajczak, Z., Nippert, J.B., Briggs, J.M. & Blair, J.M. Fire dynamics distinguish grasslands, shrublands and woodlands as alternative attractors in the Central Great Plains of North America. *J. Ecol.* **102**, 1374–1385 (2014).

18. Tamen, T. A minimalistic model of tree-grass interactions using impulsive differential equations and non-linear feedback functions on grass biomass onto fire-induced tree mortality. *Math Comput. Simul*. **133**, 265–297 (2017).

19. Ackman, O., Comar, T.D. & Hrozencik, D. On impulsive integrated pest management models with stochastic effects. *Front. Neurosci.* **9**, 10.3389/fnins.2015.00119 (2015).

20. Miller, A.D., Roxburgh, S.H. & Shea, K. How frequency and intensity shape diversity-disturbance relationships. *PNAS* **108**, 5643–5648 (2011).

21. Courchamp, F., Clutton-Brock, T. & Grenfell, B. Inverse density dependence and the Allee effect. *Trends Ecol. Evol.* **4**, 405–410 (1999).

22. Gascoigne, J. & Lipcius, R.N. Allee effects in marine systems. *Mar. Ecol. Prog. Ser.* **269**, 49–59 (2004).

23. Keith, D.M & Hutchings, J.A. Population dynamics of marine fishes at low abundance. *Can. J. Fish. Aquat. Sci.* **69**, 1150–1163 (2012).

24. Scheffer, M., Carpenter, S., Foley, J.A., Folke, C. & Walker, B. Catastrophic shifts in ecosystems. *Nature* **413**, 591–596 (2001).

25. Carpenter, S.R., Ludwig, D. & Brock, W.A. Management of eutrophication for lakes subject to potentially irreversible change. *Ecol. Appl.* **9**, 751–771 (1999).

26. Wang, R. *et al.* Flickering gives early warning signals of a critical transition to a eutrophic lake state. *Nature* **492**, 419–422 (2012).

27. Stommel, H. Thermohaline convection with two stable regimes of flow. *Tellus* **13**, 224–230 (1961).

28. Jackson, L.C., *et al*. Global and European climate impacts of a slowdown of the AMOC in a high resolution GCM. *Clim. Dyn.* **45**, 3299–3316 (2015).

29. Fairbanks, R. G. A 17,000-year glacio-eustatic sea level record: influence of glacial melting rates on the Younger Dryas event and deep-ocean circulation. *Nature* **342**, 637–642 (1989).

30. Cessi, P. A simple box model of stochastically forced thermohaline flow. *J. Phys. Oceanogr.* **24**, 1911–1920 (1994).

31. May, R.M. Thresholds and breakpoints in ecosystems with a multiplicity of stable states. *Nature* **269**, 471–477 (1977).

32. Haldane, A.G. & May, R.M. Systemic risk in banking ecosystems. *Nature* **469**, 351–355 (2011).

33. Huang, S. & Kauffman, S. How to escape the cancer attractor: rationale and limitations of multi-target drugs. *Seminars in Cancer Biology* **23**, 270–278 (2013).

34. Hirsch, M.W., Smale, S. & Devaney, R.L. *Differential Equations, Dynamical Systems, and an Introduction to Chaos*. Elsevier, Oxford (2013).

**Supplementary Information**

Here we present some mathematical arguments that underpin our work. Figure S1 portrays a general one-dimensional flow-kick system that we use to frame many of our results.

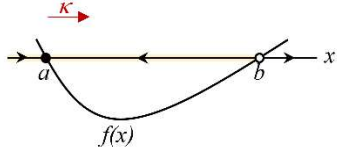

**Figure S1** A general one-dimensional flow-kick system (see text for details).

In this context, the ordinary differential equation $x' = f(x)$, where $x'$ denotes $dx/dt$, dictates the undisturbed dynamics. We assume throughout that $f$ is sufficiently differentiable (e.g. $C^1$). For a flow time $\tau$ and kick size $\kappa$, the flow-kick map is defined by $G_{\tau,\kappa}(x) = \varphi_\tau(x) + \kappa$, where $\varphi_\tau(x)$ is the flow generated by vector field $f$ [34]. We focus on the resilience of the basin of attraction, $\mathcal{B}(a)$, of the attracting equilibrium $a$ to kicks $\kappa > 0$. These kicks counteract the undisturbed dynamics, which flow for time $\tau$ toward attracting equilibrium $a$ and away from repelling equilibrium $b$ (the relevant boundary of $\mathcal{B}(a)$).

We refer to this case with $a < b$, $f(x) < 0$ on $(a,b)$, $f'(a) < 0$, $f'(b) > 0$, and $\kappa > 0$ as *the hypotheses of Figure S1*. Note that rotation of Figure S1 by 180 degrees yields a picture in which $a > b$, $f(x) > 0$ on $(a,b)$, $f'(a) < 0$, $f'(b) > 0$, and $\kappa < 0$; arguments that use the hypotheses of Figure S1 could be easily modified to handle this alternate case.

Several results below rely on the following fact.

**Lemma 1:** A one dimensional flow-kick map $G_{\tau,\kappa}$ is monotone.

**Proof:** Suppose $x < y$. Then monotonicity of flow on a one-dimensional space implies

$\varphi_\tau(x) < \varphi_\tau(y)$. Adding the same kick to both sides preserves the inequality: $\varphi_\tau(x) + \kappa < \varphi_\tau(y) + \kappa$. Hence $G_{\tau,\kappa}(x) < G_{\tau,\kappa}(y)$. //

## 1. Plotting the resilience boundary based on minimum recovery times

***Definition:*** The resilience boundary, $R$, for the basin of attraction of $a$, $\mathcal{B}(a)$, separates disturbance patterns $(\tau, \kappa)$ for which the flow-kick trajectory of $a$ remains in $\mathcal{B}(a)$ from disturbance patterns for which the flow-kick trajectory of $a$ escapes from $\mathcal{B}(a)$.

***Proposition 1:*** In addition to the hypotheses of Figure S1, suppose that $f$ is unimodal between $a$ and $b$. Then a disturbance $(\tau, \kappa)$ belongs to the resilience boundary $R$ for the basin of $a$ exactly when $\tau$ is the minimum time in which the system can recover by $-\kappa$, taken over all possible intervals of length $|\kappa|$ between $a$ and $b$.

***Proof:*** Unimodality of the vector field function guarantees a unique interval of length $|\kappa|$ over which the minimum flow time, $m$, occurs; call it $[y - \kappa, y]$. Geometrically, the interval $[y - \kappa, y]$ corresponds to 'slicing' a cap of width $\kappa$ off the bottom of the graph of $f$, since that corresponds to the fastest flow rates. If $\tau$ is greater than $m$, then a flow-kick trajectory that starts at $a$ will not escape from the basin of attraction of $a$. This is because $\tau > m$ implies $\varphi_\tau(y) < y - \kappa$, so $G_{\tau,\kappa}(y) < y$; monotonicity of the 1D flow-kick map (Lemma 1) then implies $G_{\tau,\kappa}^n(a) < G_{\tau,\kappa}^n(y) < y < b$ for all $n$. On the other hand, if $\tau$ is less than $m$, then for any $x$ in $[a, b]$ we have $\varphi_\tau(x) > x - \kappa$ so $G_{\tau,\kappa}(x) > x$. Compactness of $[a,b]$ guarantees the existence of an $\varepsilon$ such that $G_{\tau,\kappa}(x) - x > \varepsilon > 0$ for each $x$ in $[a, b]$ and so a flow-kick trajectory starting at $a$ escapes from the basin of attraction of $a$ within $j$ flow-kick iterations where $j$ is the smallest integer greater than $(b - a)/\varepsilon$. Hence $(\tau=m, \kappa)$ lies on the resilience boundary between disturbance patterns for which the flow-kick trajectory of *a* escapes from $\mathcal{B}(a)$ and disturbance patterns for which the flow-kick trajectory of *a* remains in $\mathcal{B}(a)$. //

## 2. Determining finite area of the "nonresilient" region

Recall that the nonresilient region of disturbance space, denoted *N* (see Figure 2b), is the subset of disturbance space with $\tau > 0$ that lies above the resilience boundary *R* and below the horizontal asymptote at $\kappa = |a - b|$

***Proposition 2.*** Under the hypotheses of Figure S1, the area of the nonresilient region, *N*, is finite.

***Proof:*** We first show that the region *N* has finite area when the vector field for the undisturbed system is a quadratic function $q(x)=r(x–a)(x–b)$ with $r > 0$. For any kick $\kappa$ between 0 and $b – a$, symmetry of the parabola $y=q(x)$ implies that the shortest time in which the system can recover by $–\kappa$ between *a* and *b* occurs over an interval of length $\kappa$ centered on $(a+b)/2$. Separation of variables applied to $x'=q(x)$ yields this minimum recovery time as a function of $\kappa$:

$$\tau = F(\kappa) = \int_{\frac{a+b+\kappa}{2}}^{\frac{a+b-\kappa}{2}} \frac{dx}{r(x-a)(x-b)}.$$

The resilience boundary is composed of points $(F(\kappa), \kappa)$ (see section S1). Integrating and solving for $\kappa$ in terms of $\tau$ yields a function for the resilience boundary:

$$\kappa = (b-a)\tanh\frac{r(b-a)\tau}{4}.$$

The area of the nonresilient region, *N,* is then

$$\lim_{T\to\infty} \int_0^T (b-a)\left(1 - \tanh\left(\frac{r(b-a)\tau}{4}\right)\right) d\tau$$

$$= (b-a) \lim_{T \to \infty} \left(T - \frac{\ln(\cosh(CT))}{C}\right) \quad \text{(where} \quad C = \frac{r(b-a)}{4})$$

$$= (b-a) \lim_{T \to \infty} \left(T - \frac{\ln\left(\frac{e^{CT}+e^{-CT}}{2}\right)}{C}\right)$$

$$= (b-a) \lim_{T \to \infty} \left(T - \frac{\ln\left(\frac{e^{CT}}{2}\right)}{C}\right) \quad \text{(since } e^{-CT} \to 0 \text{ as } T \to \infty)$$

$$= \frac{(b-a)\ln 2}{C}$$

$$= \frac{4 \ln 2}{r},$$

a finite quantity.

The proposition then follows from the facts that

(i) by choosing $r$ small enough, we may ensure that $0 < q(x) < f(x)$ for all $x$ in $(a,b)$, and

(ii) if $0 < q(x) < f(x)$ for all $x$ in $(a,b)$, then the nonresilient region for $x'=f(x)$ is a subset of the nonresilient region for $x'=q(x)$.  //

When approximating the finite area of a nonresilient region, fact (ii) can yield an analytic bound on the tail of the improper integral. For example, if $f(x) = x\left(1 - \frac{x}{100}\right)\left(\frac{x}{20} - 1\right)$, the quadratic function $q(x) = 20\left(1 - \frac{x}{100}\right)\left(\frac{x}{20} - 1\right)$ satisfies $0 < q(x) < f(x)$ on $(20, 100)$. As in the proof of Proposition 2, we can analytically determine both the resilience boundary for $x' = q(x)$ (here, $\kappa(\tau) = 80 \tanh(\tau/5)$ ) and the improper integral $I = \int_{\tau_0}^{\infty} (80 - 80 \tanh\frac{\tau}{5})d\tau$. Since the nonresilient region of $x' = q(x)$ is a subset of that of $x'=f(x)$, the area of the region between the resilience boundary for the cubic system and the horizontal asymptote over the tail $\tau > \tau_0$ can be no larger than $I$.

## 3. Connecting the distance-to-bifurcation resilience metric to the slope of the resilience boundary near the origin.

To understand the relationship between the slope of the resilience boundary near the origin and the distance to bifurcation in the undisturbed system $x'=f(x)$, we first need the following fact.

***Proposition 3.*** If the flow time $\tau$ and kick $\kappa$ in a flow-kick system based on an undisturbed system $x'=f(x)$ are held in a constant ratio $\kappa/\tau = r$, then, in the limit as $\tau$ goes to zero the flow-kick system limits to the continuous system $x'=f(x)+r$.

***Proof:*** At any point $x$ in state space, the flow map is $\varphi_\tau(x) = x + \tau f(x) + O(\tau^2)$, where $O$ is the big-O Landau symbol representing terms of order $\tau^2$ and higher. The flow-kick map is thus $G_{\tau,\kappa}(x) = x + \tau f(x) + O(\tau^2) + \kappa$, and since $\kappa/\tau = r$, this is $G_{\tau,\kappa}(x) = x + \tau f(x) + r\tau + O(\tau^2)$. Rearranging terms gives

$$\frac{G_{\tau,\kappa}(x) - x}{\tau} = f(x) + r + O(\tau)$$

Taking the limit as $\tau$ goes to zero yields

$$\lim_{\tau \to 0} \frac{G_{\tau,\kappa}(x) - x}{\tau} = \lim_{\tau \to 0}(f(x) + r + O(\tau))$$

The left side of this equation represents the vector field generated by the infinitesimal flow-kick map, while the right side simplifies to $f(x)+r$, as claimed. //

Now we introduce a parameter $\mu$ to the underlying system. So $x'=f(x)+\mu$, and when $\mu=0$, $f(x)$ has attracting and repelling equilibria at $a$ and $b$, respectively, as in Figure S1.

***Corollary 3.1*** If the underlying one-parameter system $x'=f(x)+\mu$ has a saddle-node bifurcation at $\mu=\mu^*>0$, at which the attracting and repelling equilibria corresponding to $a$ and $b$ coalesce and disappear, then the slope $dR/d\tau$ of the resilience boundary $R(\tau)$ converges to $\mu^*$ as $\tau$ approaches 0.

***Proof:*** Proposition 3 implies that by choosing small enough $\tau$ and $\kappa$ along the line $\kappa = r\tau$ in $(\tau, \kappa)$ space, trajectories of the flow kick system $x_{n+1} = G_{\tau,\kappa}(x_n)$ can be arbitrarily well-approximated by those of the continuous system $x' = f(x) + r$. If $\mu_1 < \mu^*$, then trajectories of flow-kick systems for $f$ with disturbance patterns approaching the origin along the line $\kappa = \mu_1 \tau$ in $(\tau, \kappa)$ space become arbitrarily close to those of $x' = f(x) + \mu_1$. This system retains an attracting equilibrium in $[a,b]$. Thus, for sufficiently small $\tau$ and $\kappa$ with $\kappa = \mu_1 \tau$, $(\tau, \kappa)$ lies below $R$ in the "resilient" region of $(\tau, \kappa)$ space. On the other hand, if $\mu_2 > \mu^*$, then trajectories of flow-kick systems approaching the origin along the line $\kappa = \mu_2 \tau$ become arbitrarily well-approximated by those of $x' = f(x) + \mu_2$. This system has lost the attracting equilibrium, and all trajectories escape from $[a,b]$. Thus, for sufficiently small $\tau$ and $\kappa$ with $\kappa = \mu_2 \tau$, $(\tau, \kappa)$ lies above $R$ in the "nonresilient" region of $(\tau, \kappa)$ space. As the borderline between these two cases, the resilience boundary $R(\tau)$ must have a slope limiting to $\mu^*$ as $\tau$ approaches zero. //

## 4. Extending deterministic results to predict outcomes of stochastic disturbances

The following propositions formalize and generalize to one dimensional systems the statements that if $\tau$ and $\kappa$ are drawn from a bounded rectangular subset either entirely below or entirely above the resilience boundary (e.g. the solid- or dashed-edged boxes in Figure 2b), then randomized flow-kick trajectories starting at the attracting equilibrium will stay within its basin of attraction indefinitely or escape in finite time, respectively.

***Proposition 4.*** In addition to the hypotheses of Figure S1, suppose that for each disturbance in the rectangle $\{(\tau,\kappa): 0 < \tau_1 \leq \tau \leq \tau_2,\ 0 < \kappa_1 \leq \kappa \leq \kappa_2\}$ the deterministic flow-kick trajectory $\{a, G_{\tau,\kappa}(a), G_{\tau,\kappa}^2(a), \ldots\}$ remains bounded in the interval $[a,b]$. Then if kicks $K$ and recovery periods $T$ vary stochastically but satisfy $\kappa_1 \leq K \leq \kappa_2$ and $\tau_1 \leq T \leq \tau_2$, any resulting flow-kick trajectory that starts at $a$ also stays within $[a,b]$.

***Proof:*** An inductive argument (below) shows that after $n \geq 0$ iterations of the flow-kick map,

$$G_{\tau_2,\kappa_1}^n(a) \leq G_{T,K}^n(a) \leq G_{\tau_1,\kappa_2}^n(a).$$

Since by assumption $a \leq G_{\tau_2,\kappa_1}^n(a)$ and $G_{\tau_1,\kappa_2}^n(a) \leq b$, we have that for all $n \geq 0$

$$a \leq G_{T,K}^n(a) \leq b,$$

as desired.

The induction proceeds on iterations $n$ of the flow-kick map. In the first flow, (because $a$ is an equilibrium of the flow) we have $\varphi_{\tau_2}(a) = \varphi_T(a) = \varphi_{\tau_1}(a) = a$, and after the first kick $\varphi_{\tau_2}(a) + \kappa_1 \leq \varphi_T(a) + K \leq \varphi_{\tau_1}(a) + \kappa_2$, establishing the base case $G_{\tau_2,\kappa_1}(a) \leq G_{T,K}(a) \leq G_{\tau_1,\kappa_2}(a)$.

Now for any $m>0$, if $G_{\tau_2,\kappa_1}^m(a) \leq G_{T,K}^m(a) \leq G_{\tau_1,\kappa_2}^m(a)$, monotonicity of the flow implies that $\varphi_{\tau_2}\left(G_{\tau_2,\kappa_1}^m(a)\right) \leq \varphi_T\left(G_{T,K}^m(a)\right) \leq \varphi_{\tau_1}\left(G_{\tau_1,\kappa_2}^m(a)\right)$, while the ordering of the kicks implies

$\varphi_{\tau_2}\left(G_{\tau_2,\kappa_1}^m(a)\right) + \kappa_1 \leq \varphi_T\left(G_{T,K}^m(a)\right) + K \leq \varphi_{\tau_1}\left(G_{\tau_1,\kappa_2}^m(a)\right) + \kappa_2$. This is equivalent to

$G_{\tau_2,\kappa_1}^{m+1}(a) \leq G_{T,K}^{m+1}(a) \leq G_{\tau_1,\kappa_2}^{m+1}(a)$, completing the inductive step. //

***Proposition 5.*** In addition to the hypotheses of Figure S1, suppose that for fixed $\tau$ and $\kappa$ the deterministic flow-kick trajectory $\{a, G_{\tau,\kappa}(a), G_{\tau,\kappa}^2(a), \ldots\}$ exceeds $b$ in finite time $t_{\tau,\kappa}$. Then if kicks $K$ and recovery periods $T$ vary stochastically but satisfy $\kappa \leq K$ and $T \leq \tau$, any resulting flow-kick trajectory that starts at $x=a$ also leaves $[a,b]$ in finite time.

*Proof:* As in the proof of Proposition 4, we have that

$$G^n_{\tau,\kappa}(a) \leq G^n_{T,K}(a).$$

Since the trajectory $\{G^n_{\tau,\kappa}(a)\}_{n=0}^{\infty}$ exceeds $b$ in finite time, so does any realized trajectory $\{G^n_{T,K}(a)\}_{n=0}^{\infty}$. //

*Corollary 5.1* If in addition to the hypotheses of Figure S1, for each $\tau$ and $\kappa$ in the rectangular domain $D = \{(\tau,\kappa) : 0 \leq \tau_1 \leq \tau \leq \tau_2, 0 \leq \kappa_1 \leq \kappa \leq \kappa_2\}$ the deterministic flow-kick trajectory $\{a, G_{\tau,\kappa}(a), G^2_{\tau,\kappa}(a), \ldots\}$ exceeds $b$ in finite time $t_{\tau,\kappa}$, then if kicks $K$ and recovery periods $T$ are drawn stochastically from $D$, any resulting flow-kick trajectory that starts at $x=a$ also leaves $[a,b]$ in finite time.

## 5. Calculating flow-kick equilibria via Newton's method

Newton's method was implemented in the MATLAB code *Newton.m* to find zeros of $F(x) = G_{\tau,\kappa}(x) - x$ for fixed values of $\tau$ and $\kappa$. The iterated step in this algorithm is

$$x_{n+1} = x_n - \frac{G_{\tau,\kappa}(x_n) - x_n}{D_x(G_{\tau,\kappa}(x) - x)|_{x=x_n}} = x_n - \frac{G_{\tau,\kappa}(x_n) - x_n}{D_x(G_{\tau,\kappa}(x))|_{x=x_n} - 1} \quad (1)$$

The term $D_x(G_{\tau,\kappa}(x))|_{x=x_n}$ in (1) simplifies to $D_x(\varphi_\tau(x))|_{x=x_n}$ because $G_{\tau,\kappa}(x) = \varphi_\tau(x) + \kappa$, with $\kappa$ constant. The spatial ($x$) derivative of the flow function $\varphi_\tau(x)$ corresponding to the vector field $f(x)$ relates closely to the solution of the variational equation

$$du/dt = [D_x(f)|_{x(t)}]\, u \quad (2)$$

which for initial condition $u=u_0$ is[1] $u(t)=D_x[\varphi_t]\, u_0$ with the derivative $D_x[\varphi_t]$ evaluated at $x=x_0$. When computing Newton iterates, it is natural to couple the evolution of $x$ by $x'=f(x)$ with the evolution of u by equation (2); the code *CoupledVar.m* contains these coupled ODEs. By numerically solving the variational equation for $u(\tau)$ with $u_0=1$ and $x_0=x_n$, we obtained the desired term $D_x(\varphi_\tau(x))|_{x=x_n}$ for each step in the algorithm (1).

## 6. Analyzing stability of flow-kick equilibria

***Proposition 6:*** Suppose $x^*$ is an equilibrium for the flow-kick map $G_{\tau,\kappa}$ in a one-dimensional system and let $f$ be the continuously differentiable vector field for the undisturbed dynamics. Let $x(t)$ represent the flow trajectory from $x^*$ to $x^* - \kappa$. If $\int_0^\tau [D_x(f)|_{x(t)}]\, dt$ is negative, then $x^*$ is a stable flow-kick equilibrium; if $\int_0^\tau [D_x(f)|_{x(t)}]\, dt$ is positive, then $x^*$ is an unstable flow-kick equilibrium.

***Proof:*** The linearization of the flow-kick map about the equilibrium $x^*$ is the map sending $y$ to $(D_x G_{\tau,\kappa}|_{x=x^*})\, y$, where $y = x - x^*$. Note that $D_x G_{\tau,\kappa}|_{x=x^*} = D_x(\varphi_\tau(x)+\kappa)|_{x=x^*} = D_x\, \varphi_\tau\, |_{x=x^*}$ since $\kappa$ is constant. As in the previous section, the spatial ($x$) derivative of the flow function $\varphi_\tau(x)$ corresponding to the vector field $f(x)$ relates closely to the solution of the variational equation

$$du/dt = [D_x(f)|_{x(t)}]\, u \qquad (2)$$

where $x(t) = \varphi_t(x_0)$ (i.e. the derivative $D_x(f)$ is computed along the trajectory from $x_0$). Namely, the solution to the variational equation (1) with initial condition $u(0) = u_0$ is [34]

$$u(t) = \left(D_x(\varphi_t)|_{x=x_0}\right) u_0\, . \qquad (3)$$

Therefore, the desired derivative $D_x\varphi_\tau|_{x=x^*}$ is precisely the solution $u(\tau)$ to the variational equation (2) at time $\tau$ with initial condition $u_0 = 1$ and x-trajectory starting at $x_0 = x^*$. In a one-dimensional system, one can solve (2) for $u(\tau)$ analytically via separation of variables:

$$\frac{du}{u} = [D_x(f)|_{x(t)}]\, dt$$

$$\int_{u(0)=1}^{u(\tau)} \frac{du}{u} = \int_0^\tau [D_x(f)|_{x(t)}]\, dt$$

$$\ln(u(\tau)) - \ln(u(0)) = \int_0^\tau [D_x(f)|_{x(t)}]\, dt$$

$$u(\tau) = \exp\left(\int_0^\tau [D_x(f)|_{x(t)}]\, dt\right).$$

If $\int_0^\tau [D_x(f)|_{x(t)}]\, dt$ is negative, then $u(\tau) = D_x\varphi_\tau|_{x=x^*}$ is less than one, so the linearized flow-kick map has eigenvalue less that one, and $x^*$ is a stable flow-kick equilibrium; if $\int_0^\tau [D_x(f)|_{x(t)}]\, dt$ is positive, then $u(\tau) = D_x\varphi_\tau|_{x=x^*}$ is greater than one and $x^*$ is an unstable flow-kick equilibrium.   //

***Corollary 6.1:*** In the special case that $D_x(f)$ is negative (positive)—or equivalently, the potential function $V = -\int f\, dx$ is concave up (down)—over the entire flow-kick equilibrium interval, the sign of the integral $\int_0^\tau [D_x(f)|_{x(t)}]\, dt$ and hence the stability (instability) of $x^*$ can be determined by inspection.

## 7. Viewing the resilience boundary as a bifurcation curve

So far, we have calculated resilience boundaries in terms of minimum recovery times (see section 1). However, the resilience boundaries of the fishery and lake water quality examples can also be viewed as bifurcation curves at which stable and unstable flow-kick equilibrium intervals coalesce and disappear.

Figure S2 illustrates this interpretation for the lake water quality example with undisturbed dynamics $\frac{dx}{dt} = l - sx + \frac{rx^q}{m^q+x^q}$. Figure S2a shows that when $\kappa=10$ and $\tau=2$, there are three flow-kick equilibrium intervals: $A$ and $C$ (stable), and $B$ (unstable). The bifurcation diagram in Figure S2b shows how the existence and positions of the flow-kick equilibria vary with $\kappa$ while $\tau=2$ remains fixed. For each value of $\kappa$, a vertical slice shows the locations of flow-kick equilibrium intervals, bounded by the grey curve below and the black curve above. The particular case $\kappa=10$ is highlighted as an example. As $\kappa$ approaches a critical magnitude $\kappa^* \approx 15$, the stable and unstable low-P equilibrium intervals overlap, and at $\kappa = \kappa^*$ they coalesce and disappear in a saddle-node bifurcation. For $\kappa > \kappa^*$, there is a single, globally attracting equilibrium interval at high P. If $x$ starts near the deterministic equilibrium $a$, a succession of kicks that occurs every 2 time units will escape the basin of attraction of $a$ if $\kappa > \kappa^*$, or will stay in the basin if $\kappa > \kappa^*$. The pair $(2, \kappa^*)$ thus belongs on the resilience boundary for this flow-kick system.

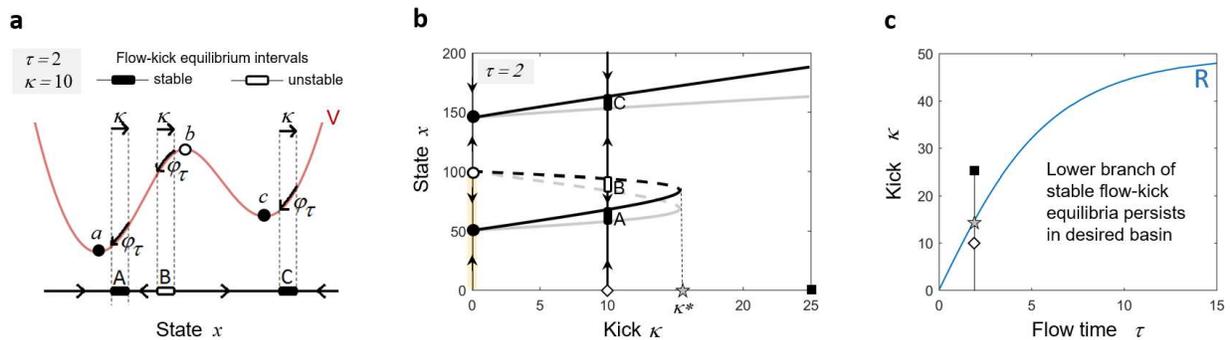

**Figure S2** The resilience curve as a bifurcation diagram. (see text for details).

Figure S2c generalizes a step further by allowing the recovery time $\tau$ to vary (horizontal axis) and plotting (solid curve $R$) the kick size $\kappa$ (now on the vertical axis) at which the saddle-node bifurcation occurs. The vertical line at $\tau = 2$ corresponds to the horizontal axis of the bifurcation diagram in Figure S2b. The curve $R$ of bifurcation values separates disturbance patterns ($\tau, \kappa$) that maintain the structure of two, alternative stable equilibrium intervals (below) from disturbance patterns for which the lower stable equilibrium interval is lost (above). $R$ is precisely our resilience boundary in disturbance space for the basin of the underlying deterministic attractor $a$.

The flow-kick equilibria in Figure S2a were calculated using Newton's method (MATLAB codes *Newton.m* and *CoupledVar.m*, section 5); their stability can be deduced from Corollary 6.1 (section 6). The branches of equilibria in Figure S2b were calculated using the MATLAB script *Branch.m*, which calls *Newton.m*.